# User-level DSM System for Modern High-Performance Interconnection Networks

Bharath Ramesh and Srinidhi Varadarajan
Computing Systems Research Lab
Department of Computer Science
Virginia Tech, VA 24061
Email: {bramesh,srinidhi}@cs.vt.edu

**Abstract**

In this paper, we introduce a new user-level DSM system which has the ability to directly interact with underlying interconnection networks. The DSM system provides the application programmer a flexible API to program parallel applications either using shared memory semantics over physically distributed memory or to use an efficient remote memory demand paging technique. We also introduce a new time slice based memory consistency protocol which is used by the DSM system. We present preliminary results from our implementation on a small Opteron Linux cluster interconnected over Myrinet.

## 1  Introduction

The bi-annual rankings of the "Top 500" supercomputers [1] over the past few years shows a major shift in High Performance Computing from custom parallel architectures and vector supercomputers towards clusters. This shift affects how inter-process communication works – the two basic models of inter-process communication being shared memory and message passing. In the case of clusters, since there is no single global physical memory, one cannot take advantage of the ease of inter-process communication provided by shared memory architectures. When using a message passing programming model, the programmer needs to take care of data distribution across the system apart from managing communication, which adds a significant burden. Distributed Shared Memory (DSM) systems provide shared memory semantics over physically distributed memory. These systems hide the underlying communication mechanism from the programmer, which is an advantage as the programming model followed can be much simpler.

Concomitantly, applications in scientific computing, engineering simulations, multimedia, etc. require an ever increasing amount of memory, usually more than what a single workstation has available. Operating systems use demand paging techniques to alleviate this memory shortage. Traditionally disks have been used as paging devices, but with the ever increasing gap between the processors and disk speeds, the cost of paging to disks has become unacceptable. This requires the design of either new paging devices or techniques to mitigate this problem. Remote memory demand paging is one such solution, in which the physical memory of another node is used as the paging device. In remote memory demand paging, the bandwidth and latency are mainly dictated by the interconnection network. Modern high-performance interconnection networks have the ability to provide low latency, high bandwidth remote memory access.

In this paper we present a new user-level DSM system for emerging networking technology and modern day multi-core multi-processor clusters. The DSM system is implemented in user-level and requires no modification to the operating system kernel. The run time system provides the programmer with a flexible Application Programming Interface (API) to use shared memory semantics over physically distributed memory or to use a remote memory demand paging technique as required by the application. Since the run time system is aware of the interconnection network, it interacts directly with the network hardware, bypassing the operating system to decrease communication latencies. The rest of the paper is divided into four sections. The first section provides brief background information about related work. In the second section we present the design of our new user-level DSM system. The results presented in the third section, though preliminary, are encouraging. The last section focuses on conclusions and future work.

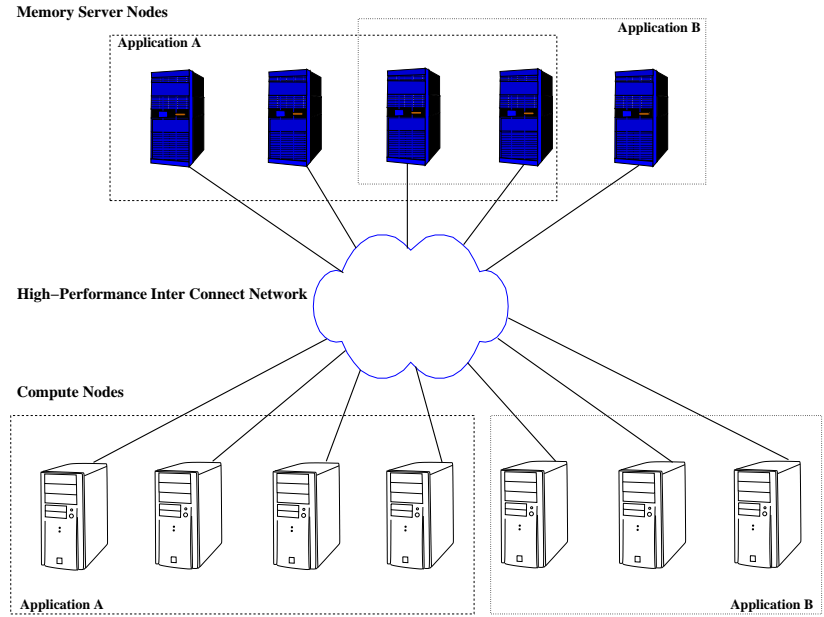

Figure 1: Typical setup of DSM system on a cluster

## 2 Related work

There have been many software DSM systems since the first system IVY [8] emerged. DSM systems like Munin [3], Midway [2], TreadMarks [7], and JIAJIA [6] were all implemented either on top of Ethernet or ATM LANs. These systems depend on the operating system to interact with the interconnection network thus increasing their communication overhead. JUMP-DP [4] was the first system to focus on reducing the communication overhead by implementing a low-latency socket model in the kernel. In our implementation we reduce the communication latency by directly interacting with the interconnection network using softwares specific for the interconnection network. Other systems which have implemented remote memory demand paging technique include Remote Memory Pager [9], Global Memory System [5], and Parallel Network RAM [10], but none of them provide the ability to use shared memory semantics over physically distributed memory.

## 3 Design of the user-level DSM system

In this section we provide a brief architectural overview of the user-level DSM system, without discussing the implementation details. We also introduce a new time slice based memory consistency model using write-invalidate as the coherency protocol.

### 3.1 Architecture

The DSM system consists of two types of nodes – memory server nodes and compute nodes. The memory servers serve the memory requirements of an application, while the computation is handled by the compute nodes in the cluster. The memory server nodes are not linked to the application running on the compute nodes. These nodes have the ability to serve not just one application but multiple applications running on the cluster utilizing our DSM system. Figure 1 shows a typical setup of applications using our DSM system. The memory server nodes have low computational requirements in our design, hence they have the ability to coexist with the compute nodes on the same physical node of the cluster. One key feature of the DSM system is the ability to directly interact with the interconnection network to reduce communication latency. In order to support a variety of interconnection networks we have followed a layered design. A node in the DSM system comprises of three basic layers – management layer, communication layer, and the interconnection network layer.

The management layer consists of three managers needed to implement various tasks of the DSM system – memory manager, barrier manager, and mutex manager. The implementations of these managers depends on the type of the node. The memory manager in the compute node is responsible for managing the access to the global address space shared by all the compute nodes of the application. The compute node has a local cache which has the copies of the DSM pages that the application is currently accessing. This cache is kept consistent using a time slice based consistency model, described in section 3.2. The memory manager present on the memory server node is responsible for the global address space addressed by any application. The global address space which every compute node accesses is divided into pages. Every page of the DSM global address space is associated with an owner memory server node. This node maintains all necessary information for the page. The information stored includes a list of readers and writers, status of the page, and compute node with the latest copy if needed. The different page statuses are – current, current on compute node, and write locked.

Synchronization is one of the most important aspects of any DSM system; it is required to prevent unexpected race conditions. Our DSM system provides the application programmer with two synchronization primitives – barriers and locks. The barrier manager and the mutex manager implement these features in our system.

## 3.2 Time slice based memory consistency protocol

Whenever a compute node requests access to a DSM page it can result in one of four scenarios:

1. Node requests read access to a page, which is current on the memory server. This is the simplest scenario in which the memory server sends a copy of the page to the requesting node.

2. Node requests read access to a page, which is current on another compute node. The memory server redirects the requesting node to the node with the most recent copy of the page.

3. Node requests write access to a page, which other compute nodes have access with read privileges. Memory server grants the requesting node write access, thereby invalidating all nodes with a copy of this page.

4. Node requests write access to a page, currently write locked by another compute node. The requesting node is redirected to the node currently holding the write lock. The other node relinquishes its read/write privileges, and transfers the privileges to the requesting node along with the current copy of the page.

When multiple writers concurrently write to the same DSM page, the result is a performance degradation of the application. Concurrent writes cause the system to be flooded with invalidation messages, which results in cache thrashing. Little or no progress is made by the writer as it can be invalidated upon just writing one or no unit of data to the DSM page. For the writers to make significant progress, and for fairness we employ a time slice based memory consistency protocol. This protocol guarantees the writer a minimum amount of time, write time slice, during which the writer need not relinquish its privileges. At the end of this time slice the writer needs to relinquish its privileges if and only if either there are other writers waiting in the write queue for the same page or another node requests the current copy of the page. The nodes need not synchronize their clocks, the end of the time slice is calculated based on the local clock. This helps us to improve the performance of the consistency protocol significantly.

# 4 Performance results

In this section we present some of our preliminary results. The experimental testbed consists of a four node cluster interconnected over Myrinet-2000 with Myrinet Express MX-2G communication software. Each node is a dual AMD Opteron 240 processor system with 1GB of physical memory. We designed a synthetic benchmark to test the performance of our DSM system. The parameters that we measure in our benchmark are the DSM systems raw memory bandwidth and access latency.

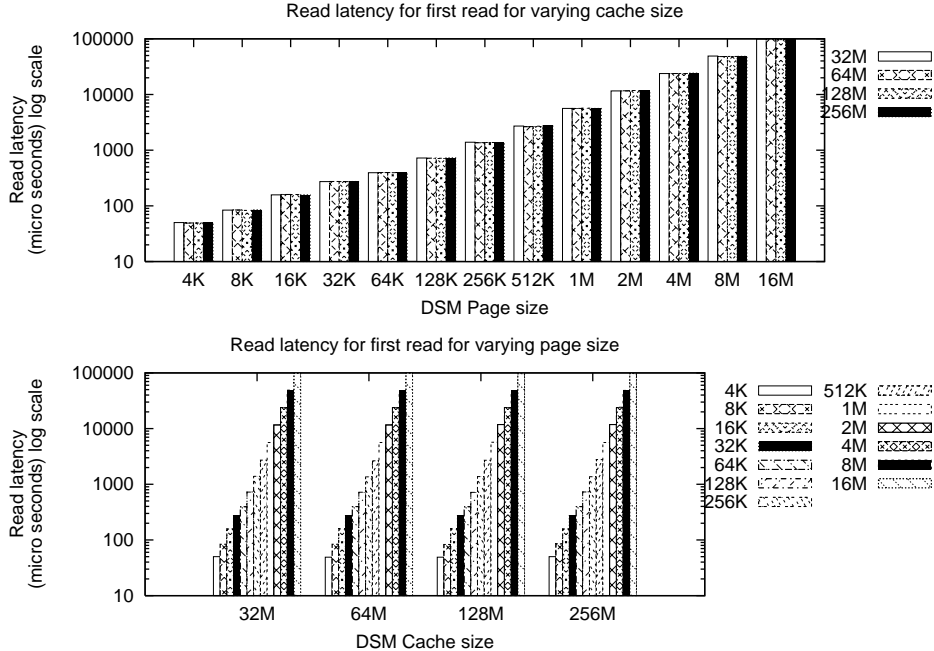

Figure 2: Read latency for varying cache and page sizes.

For our bandwidth test we used a global address space size of 256M. We measure the write bandwidth by performing a *memset* of *zero* over the entire global address space. The read bandwidth is measured by performing a *memcpy* from DSM global address space to a locally allocated memory of the same size. The access latency is measured by timing how long it takes to perform a write to the first byte of a DSM page for write latency, and reading the first byte of a DSM page for read latency. To study the effect of varying page and cache size we ran the benchmark on a wide range of DSM page size from 4K to 16M. The DSM cache size used in the benchmark were 32M, 64M, 128M and 256M.

Figures 2 and 3 show the variation in read and write latency for different page and cache sizes of the DSM system. We present the latency on the *Y-axis* using a log scale to better understand the variation in the latency for varying page and cache sizes. The latency for either read or write access solely depends only on the DSM page size. The read/write latency consists of the communication latency incurred because of the DSM system and the local memory access time.

Figures 4 and 5 show the variation in read and write bandwidth for varying page and cache sizes of the DSM system. We notice that the read bandwidth for cache size of 256M is nearly equal to that of the local memory read bandwidth since once the entire global address space is cached we only incur the local read costs. The decrease in read and write bandwidth when the page size increases from 8M to 16M can be attributed to the fact that the difference in the latencies for page size of 8M and 16M is significant and there is no gain by increasing the page size.

## 5 Conclusions and future work

In this paper we presented a user-level DSM system designed to take advantage of modern high performance interconnection networks, which uses a new time slice based memory consistency model. The design of this DSM system makes it flexible for a programmer to program parallel applications either using shared memory semantics or to exploit an efficient remote memory demand paging technique. We feel there is plenty of scope for further improving the performance of our DSM system based on the results obtained.

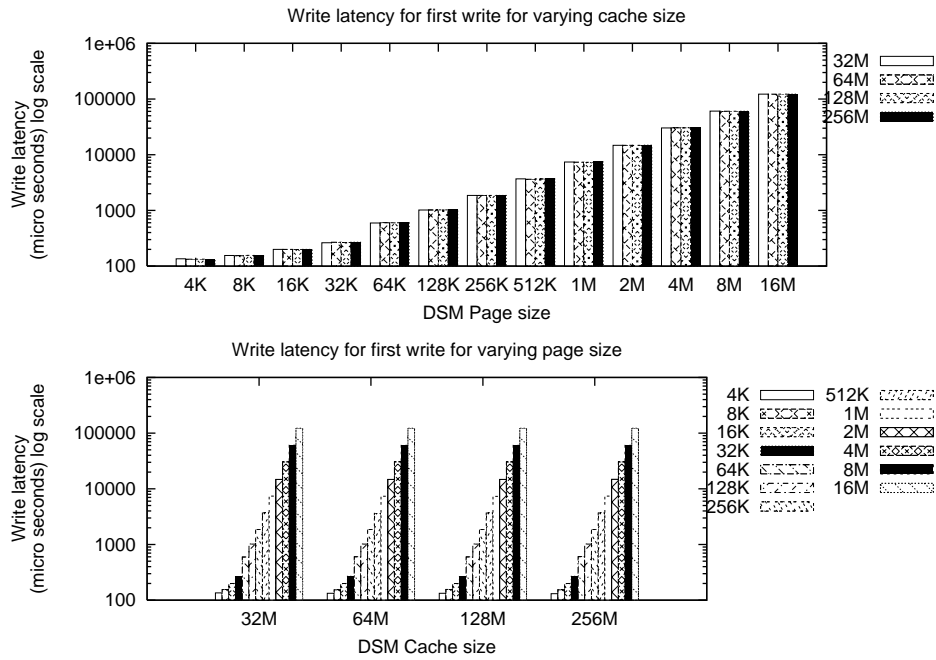

Figure 3: Write latency for varying cache and page sizes.

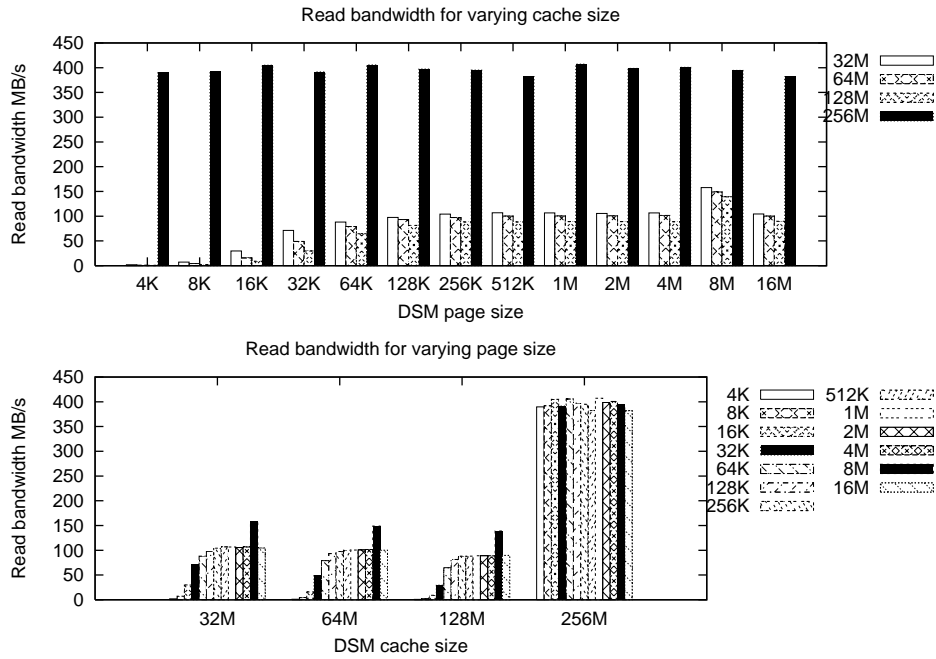

Figure 4: Read bandwidth for varying cache and page sizes.

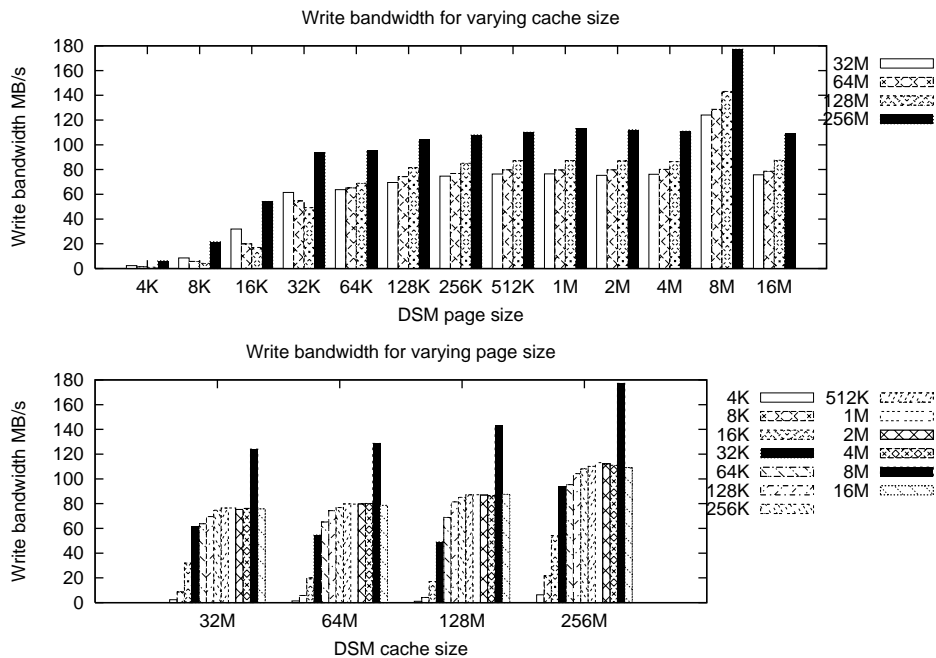

Figure 5: Write bandwidth for varying cache and page sizes.

The immediate focus is to significantly reduce the software overhead, to improve bandwidth and reduce latency. The performance of the system can be further improved by incorporating prefetch strategies to the cache management system. Finally, an efficient strategy will be devised to transfer page data, if the DSM page size is above a certain threshold, to further improve the memory bandwidth.

# References

[1] TOP500 supercomputer sites. http://www.top500.org.

[2] BERSHAD, B. N., AND ZEKAUSKAS, M. J. Midway: Shared memory parallel programming with entry consistency for distributed mrmory multiprocessors. Tech. Rep. CMU-CS-91-170, 1991.

[3] CARTER, J. B., BENNETT, J. K., AND ZWAENEPOEL, W. Implementation and performance of Munin. In *Proceedings of the 13th ACM Symposium on Operating Systems Principles (SOSP)* (1991).

[4] CHEUNG, B. W., WANG, C., AND HWANG, K. JUMP-DP: A software DSM system with low-latency communication support. In *2000 International Workshop on Cluster Computing - Technologies, Environments and Applications (CC-TEA)* (2000).

[5] FEELEY, M. J., MORGAN, W. E., PIGHIN, F. H., KARLIN, A. R., LEVY, H. M., AND THEKKATH, C. A. Implementing global memory management in a workstation cluster. In *Proceedings of the 15th ACM Symposium on Operating Systems Principles (SOSP)* (1995).

[6] HU, W., SHI, W., AND TANG, Z. JIAJIA: A software DSM system based on a new cache coherence protocol. In *High Performance Computing and Networking (HPCN) Europe* (1999).

[7] KELHER, P., COX, A. L., DWARKADAS, S., AND ZWAENEPOEL, W. Treadmarks: Distributed shared memory on standard workstations and operating systems. In *Proceedings of the Winter 1994 USENIX Conference* (1994).

[8] LI, K. IVY: A shared virtual memory system for parallel computing. In *Proceedings of the 1988 International Conference on Parallel Processing (ICPP)* (1988).

[9] MARKATOS, E. P., AND DRAMITINOS, G. Implementation of a reliable remote memory pager. In *USENIX Annual Technical Conference* (1996).

[10] OLESZKIEWICZ, J., XIAO, L., AND LIU, Y. Parallel network RAM: Effectively utilizing global cluster memory for large data-intensive parallel programs. In *2004 International Conference on Parallel Processing (ICPP)* (1994).